# Scaling limits of graphene nanoelectrodes

Syed Ghazi Sarwat, Pascal Gehring, Gerardo Rodriguez Hernandez, Jamie H. Warner,
G. Andrew. D. Briggs, Jan A. Mol & Harish Bhaskaran

Department of Materials, University of Oxford, Oxford, OX1 3PH, UK

*Graphene is an ideal material for fabricating atomically thin nanometre-spaced electrodes. Recently, carbon-based nanoelectrodes have been employed to create single-molecule transistors and phase change memory devices. In spite of the significant recent interest in their use in a range of nanoscale devices from phase change memories to molecular electronics, the operating and scaling limits of these electrodes are completely unknown. In this paper, we report on our observations of consistent voltage driven resistance switching in sub-5 nm graphene nanogaps. We find that we are able to reversibly cycle between a low and a high resistance state using feedback-controlled voltage ramps. We attribute this unexplained switching in the gap to the formation and breakdown of carbon filaments. By increasing the gap, we find that such intrinsic resistance switching of graphene nanogaps imposes a scaling limit of ~10 nm on the gap-size for devices with operating voltages of 1–2 volts.*

**Introduction** The ability to create nanometre-sized gaps in $sp^2$-bonded carbon materials offers a means of contacting nanoscale objects – for example nanocrystals and single molecules – that cannot be achieved with conventional metallic electrodes. The fact that these materials have a thickness of only a single or few atomic bond-lengths strongly reduces electrostatic screening and enables gating of molecular orbitals.[1] Moreover, the reduced contact area between atomically thin electrodes and phase change material nanocrystals has been shown to lower the power requirements for current-induced phase changes.[2] Due to the strength of the $sp^2$ carbon-carbon bond, the atomic mobility of carbon atoms is significantly lower than that of metal atoms, and carbon-based electrodes are therefore expected to be significantly more robust, even at room temperature.[3] However, we find that the intense electric fields generated by applying a bias voltage across a nanometre-size graphene gap result in the spontaneous rearrangement of atoms and bonds that lead to reversible switching of the resistance. Here, we investigate the scaling limits imposed by this switching behaviour in the context of phase change memory (PCM) devices. However, our findings carry equal significance for all applications based on graphene nanogaps, including single-molecule electronics[1,4,5] and graphene-based genome sequencing.[6]

The energy consumption and access speed of phase change memories[2] and other data storage technologies, including oxide memory,[7,8] have been shown to improve significantly as a result of scaling down the dimensions between the contact electrodes. Ultimately, the performance of these memory devices is determined by the active volume that switches between two states of contrasting electrical resistance. In theory this volume could be scaled to the dimension of a single unit cell volume[9] which requires sub-2 nm spaced electrodes. In this paper, we find that it is the intrinsic switching behaviour of the graphene electrodes, rather than the properties of the phase change material, that ultimately limits the device scaling and therefore its performance.

We use a method of feedback-controlled electroburning to create graphene nanogaps ranging from ~1 to 60 nm and, using a self-alignment approach, we deposit a small volume of $Ge_2Sb_2Te_5$ (GST) over the gap. Only in the case of large nanogaps (>20 nm) do we find that the resistance switching is due to the GST, while for smaller gaps it is fully dominated by the graphene. We characterise the graphene switching by studying bare graphene nanogaps and estimate the critical electric field for switching $F_{crit}$ = 40 mV/Å. This critical field dictates the maximum operating voltage for a given gap-size – or minimum gap-size for a given operating voltage – for any technology based on graphene nano-electrodes.

**Nanogap device fabrication**

We use a feedback-controlled electroburning[1,10] technique that relies on controlled Joule heating to form a nanoscale gap between two electrodes in an appropriately patterned graphene ribbon. This method has previously been used to create sub-5 nm gaps in mechanically exfoliated graphene[1], chemical vapour deposition (CVD) grown graphene[10,11], and epitaxial graphene[12]. Here, we use this method to create nanogaps in 2-3 layer CVD-grown graphene that was transferred onto a Si/ 300nm $SiO_2$ substrate, with Au connectors and bond pads pre-fabricated on the substrate. We use few-layered graphene rather than single-layer graphene in order to limit the effects of defects induced by sputter deposition of GST.[13] The graphene was patterned into a bow-tie geometry with a 100 nm wide constriction (see Fig. 1b) using electron beam lithography and oxygen-plasma etching. During the electro burning process, nanogaps form at the constriction, where the current density and therefore the Joule heating are highest.[10] At each stage of the electroburning process, we monitor the source-drain current when the voltage across the device is ramped up (see Fig. 1c). As the current drops, due to electroburning of graphene at the constriction, the resistance increases; the feedback-control is programmed to then ramp down the applied bias voltage back to zero. This process is repeated until the device has a resistance > 500 MΩ. By adjusting the feedback-control parameters we can fabricate nanogaps ranging from approximately 1 nm to 100 nm.

We estimate the size of the nanogaps by fitting the measured current-voltage curve to the Simmons model.[14] From these fits we find that the smallest gaps range from 0.5 nm to 3.5 nm. Using Atomic Force Microscopy (AFM) we confirm that the nanogap formation starts at the corners of the constriction and then propagates inwards (see Fig. 1b). In approximately half of the devices, we observe a sharp increase in the conductance prior to the formation of a nanogap (see inset Fig. 1c). Similar conductance enhancement behaviour has been reported before,[15–17] and is attributed to the formation of carbon filaments. Density functional theory and tight-binding simulations have shown that the transition from a multi-path configuration to a single-path configuration may lead to an enhancement of quantum transport.[17] In the following section we describe the observation of reversible resistance switching in our devices, which we attribute to the controlled formation of carbon filaments.

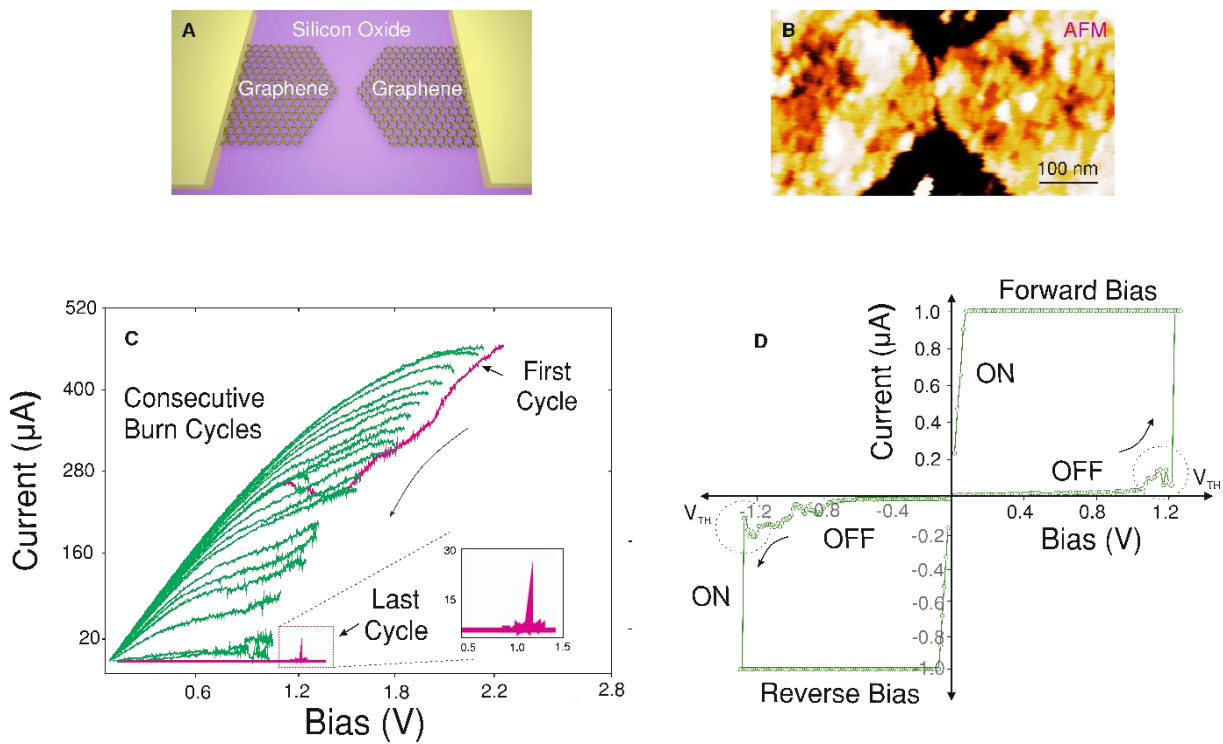

**Figure 1| A graphene nanogap device (a)** Schematic representation of a graphene nanogap device; gap size is exaggerated for visualization **(b)** AFM image of a graphene nanogap device; the gap (~ 1 nm) is not resolvable near the centre of the constriction. **(c)** Current-voltage (*I-V*) characteristics during feedback-controlled electroburning of graphene in ambient conditions. Inset represents the last cycle of the burning process, which shows a spike in conductance just before the gap forms. This current spike is attributed to single carbon filament formation. **(d)** Low bias switching of a graphene nanogap device (3 nm gap size) in ambient conditions. The first quadrant represents switching with a forward (positive) bias. The device switched from a high resistive state to a low resistive state in ambient conditions at a switching voltage of 1.22 V, and current 60 nA. The third quadrant shows the current-voltage characteristics of the same device under reverse (negative) polarity. The device switched at a voltage of 1.28 V, and current of 100 nA. For all reversible switching experiments, a current

compliance of 1000 nA was used.

**Reversible switching in graphene nanogaps**

After we form a nanogap using feedback-controlled electroburning, we can set the device back to its low resistance state by sweeping the bias voltage past a threshold voltage in ambient conditions (see Fig. 1d). We observe that this switching behaviour is independent of the bias polarity, after having switched the device from the high resistive 'OFF' to the low resistive 'ON' state by applying a forward bias we switch it OFF by repeating the electroburning and then switch it ON again by applying a negative bias. As shown in Figure 2b, the conductance switching is fully reversible; we can switch the device from the ON to the OFF state by performing the feedback-controlled electroburning process (see Fig. 2c); and switch back from the OFF to the ON state by sweeping the bias voltage beyond the threshold voltage (Fig. 2d). We can repeat SET (from OFF to ON) and RESET (from ON to OFF) multiple times.

Reversible conductance switching of graphene nanogaps has previously been reported for graphene on $SiO_2$ and suspended graphene in vacuum[18–20]. The temperature dependence observed in these studies, as well as in this paper provides a strong indication that the switching process involves the rearrangement of atoms and/or chemical bonds that requires overcoming a barrier[18]. A possible mechanism for this rearrangement is the formation of carbon filaments, which in the case of carbon nanotubes was identified as the process through which they unravel by the action of an electric field.[21] Figure 2a shows a schematic depiction of the filamentation process; the force exerted by the electric field breaks the C-C bond of an edge atom with incomplete $sp^2$ bonding. The filamentation process then proceeds as a rupture of C-C bonds parallel to the graphene edge. The fact that we observe reversible switching in ambient conditions is potentially because of the feedback-control when switching the device OFF. The gap size resulting from electroburning without feedback-control strongly depends on the oxygen concentration of the atmosphere, and ranges between ~100 nm in ambient condition to ~5 nm under a vacuum ~$10^{-5}$ mbar.[22] We find that we are unable to SET devices when electroburning without feedback-control.

The SET requires a field strength $F_{crit}$ = 40 mV/Å by assuming to a first approximation that the applied bias voltage drops linearly across the 0.75 nm gap. This field strength is similar to that observed previously[19] for a gap size of ~ 10 nm, which switched at ~4 V, suggesting that there is a critical field strength required to unzip the carbon filament(s) from graphene. Interestingly, this electric field strength is two orders of magnitude lower than the field strength that has been theoretically estimated ( ≥2 V/Å)[23,24] for unravelling a carbon filament from a graphene edge. We attribute this discrepancy to weakening of the C-C bond strength resulting from incomplete sp$^2$ hybridization and enhancement of the local electric field at atomically sharp graphene edges.[25]

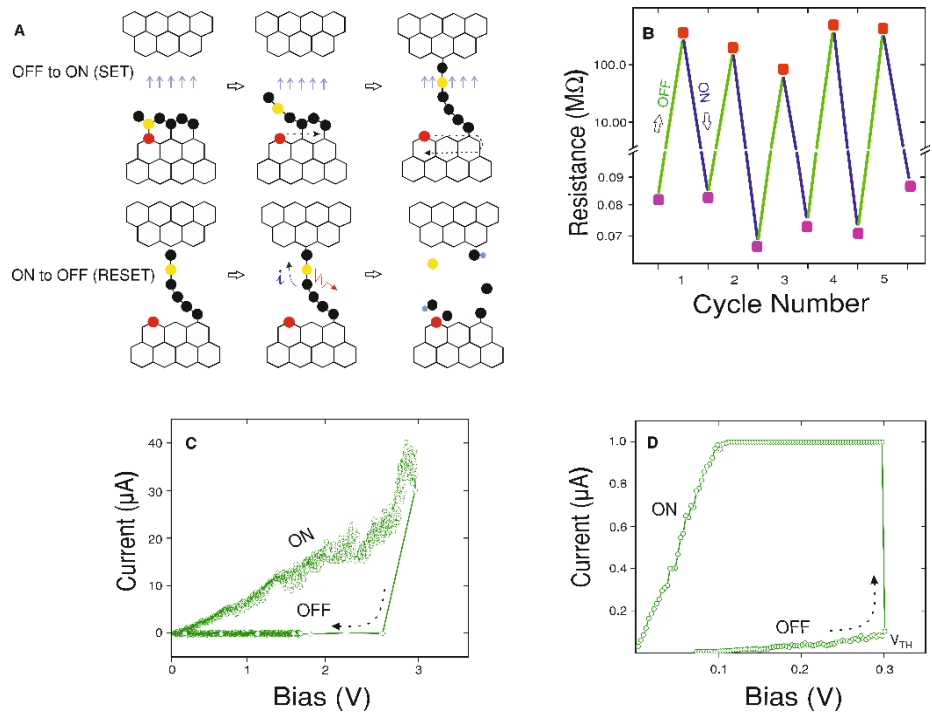

**Figure 2| Cyclic switching via filamentation in a graphene nanogap device. (a)** Proposed scheme for the formation and breaking of carbon filaments following Ref. 21: During SET (OFF to ON), formation of a carbon filament initiates from the edge of the graphene; when the local electric field at atomically sharp edges gets sufficiently high, it breaks a bond parallel to the axial electric field (between red and yellow atoms in the schematic). Filamentation then proceeds in a row-by-row fashion as indicated by the dashed arrows. For RESET (ON to OFF), Joule heating provides sufficient thermal energy for the rupture of bonds through oxidation of carbon atoms (oxygen represented by blue circles). **(b)** The device is switched between the high resistance and low resistance state multiple times in ambient conditions. **(c)** Current voltage behaviour during RESET in ambient conditions showing similarity to electroburning traces in the previous electroburning cycle (Figure 1c). **(d)** Illustrates a typical SET *I-V* characteristic in ambient conditions. The device switched at a switching voltage of 300 mV, and current 80 nA.

**Switching graphene nanogaps with GST**

Based on measurements of the critical field required for switching graphene, we estimate that to switch a Ge$_2$Sb$_2$Te$_5$ (GST) volume with a voltage less than 4 V, we require a gap size of at least 10 nm. To demonstrate this, we compare GST contacted in both 1 nm and 20 nm wide graphene nanogaps. To place the GST volume over the graphene nanogap, we use a self-alignment method that relies on the local removal of PMMA in the vicinity of the graphene constriction during the electroburning process. Similar self-alignment techniques have been previously demonstrated for fabrication of CNT nanogaps based PCM devices[26], however not in combination with feedback-controlled electroburning. After several cycles of electroburning, we spin-coated ~100 nm of poly methyl methacrylate (PMMA) onto our devices. Continuing the feedback-controlled electroburning process, we locally heat up the graphene constriction, which leads to the formation of trenches resulting from the local evaporation of PMMA. These trenches serve as self-aligned windows for subsequent deposition of the phase change material, which in our demonstrator case is GST. The size of the trenches depends on the number of electroburning cycles, i.e. the resistance of the graphene device, prior to spinning the resist. We have simulated the electroburning process using finite element analysis. The resulting trench sizes agree well with our experimental observations. Figure 3a and b show an AFM image of a self-aligned trench in PMMA, and a SEM image of the device after sputter-deposition of GST (~ 12 nm) and PMMA lift-off. We avoided capping layers in order to eliminate any probable interfacial interactions between the capping layer and GST, which are known to influence switching behaviour[27].

Figure 3c shows the current-voltage characteristics of a self-aligned PCM device in a 1 nm wide nanogap. The device switches from a highly resistive state to a low resistive state at ~ 2.5 V (Fig 3 d) similar to the observed switching in bare graphene nanogaps in Figure 2a and b. GST is a semiconductor in both its amorphous and crystalline states, and we therefore do not expect to observe a linear *I-V* character in either the ON and the OFF state. However, the *I-V* characteristics of the ON state of the GST in a 1 nm gap is linear, similar to the bare nanogap. From this, we infer that the switching in the 1 nm nanogap is dominated by the formation of carbon filaments. By contrast, for GST deposited in a 20 nm gap, the *I-V* characteristics shows an exponential dependence in both the ON and the OFF state, in agreement with previous measurements of GST. Switching in GST devices occurs at a relatively high power and the resistance ratio between the highly resistive and less resistive state is ~1000, indicative of switching in GST[4,15]. Finally, we test a device with a ~20 nm nanogap without GST and find that the device has an open circuit characteristics, displaying no switching behaviour even at very large bias values. At very large voltages of ~120-150 V, dielectric breakdown of the underlying $SiO_2$ substrate is seen to occur. We attribute the absence of filamentation in wider gaps to instabilities of long carbon filaments[28,29].

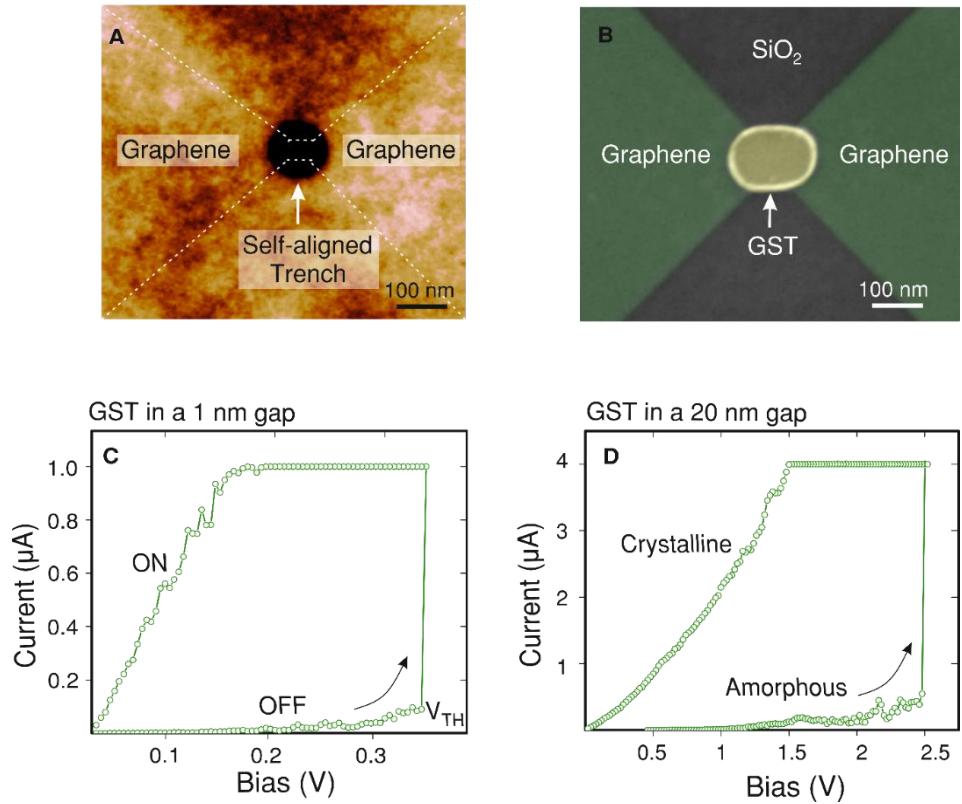

**Figure 3| Self-alignment approach and Phase Change Memory device (a)** AFM image showing a trench of size 148 nm (largest lateral dimension) in PMMA. This trench is formed in-situ from local degradation of PMMA due to Joule heating during the electroburning process. Dotted line outlines the graphene ribbon underneath PMMA. **(b)** Coloured SEM image of a self-aligned PCM device showing the phase change material (GST) in the nanogap. The trench in the PMMA ensures that the GST is self-aligned to the gap in the graphene electrodes, thus eliminating the need for sub-10nm alignment. **(c)** Current-voltage trace of a GST nanogap device; GST is aligned to make contact to graphene in a 1 nm nanogap. The device switches from a high resistive state to a low resistive state in ambient conditions at a switching voltage of 370 mV, and current 100 nA. **(d)** Current-voltage characteristics of a GST device with a gap size of ~ 20 nm. GST switches from a highly resistive amorphous state to a less resistive crystalline state at a bias of 2.5 V and current 500 nA. The ratio between these states averaged to ~ 1000.

**Further Experiments and Discussion**

Reversible conductance switching has also been observed in $SiO_2$-based devices. To exclude effects[7,8] of $SiO_2$ mediated conductance switching we carried-out two experiments. In the first experiment, we placed 15 nm thick $SiO_2$ in the sub-4 nm gaps using the self-alignment technique. We observed no switching behaviour, other than dielectric breakdown at ~ 10 V. In the second, we created graphene nano-gaps on an SiN substrate, a material that shows no intrinsic switching[8]. We observed a similar switching behaviour on this substrate as observed on the $SiO_2$ substrate. Furthermore, formation of Si nanoclusters through reduction of $SiO_2$ is recognized as the mechanism behind resistance switching in $SiO_2$ switching. Therefore an oxygen deficient atmosphere is a prerequisite[13,36-37] for switching in unpassivated $SiO_2$. Our devices can be switched both ways readily in ambient conditions. It is therefore highly unlikely that $SiO_2$ switches in our devices since the switching site, which is the surface, is exposed to an oxygen rich atmosphere. In addition, the ratio of resistance between the OFF and the ON state is typically[7,8,30] $> 10^4$ in $SiO_2$, which is ten orders magnitude more than observed in our devices

Having thus established sufficient evidence for switching from carbon filament(s) formation in nanogaps, an important question is how filamentation is possible when the phase change materials we use ($Ge_2Sb_2Te_5$ or GST) fills the 0.75 nm gap. The answer lies in structuring of the GST film during sputter deposition. Chalcogenide (which GST is) atoms show strong bonding preference for each other over $SiO_2$ for reasons relating to minimization of strain and surface energies and in the case of GST on $SiO_2$, this results in poor adhesion with the $SiO_2$ substrate[31]. Thus, it is expected that the island growth mode or the Volmer-Weber mode is preferred over layer by layer growth mode during deposition[32]. Furthermore, graphene shows a catalytic property towards the growth of chalcogenides[33]. This would result in the GST islands on graphene growing in all directions; bridging, but not filling the gap. This is supported by the absence of switching in the graphene nano-gaps with $SiO_2$ in the gap. Thus, there is a strong suggestion that regardless of the switching mechanism, there is a fundamental limit to scaling graphene nano-gaps for such relevant material systems. This perhaps also applies for carbon nanotube nano-gaps, which share similar bonding configuration ($sp^2$) as graphene, and could be a subject of future work. Importantly, molecular electronics where the actual gap is not filled entirely by the molecule, but has several areas where such chains can grow, might also have a similar scaling limit.

Therefore, our observations strongly point towards resistance switching in graphene nanogaps, which we attribute to the controlled formation and breakdown of carbon filaments. Analysing the switching behaviour, we find that the formation of carbon filaments is electric field dependent and only occurs in sub-5 nm gaps. These experiments demonstrate for the first time, reversible resistance switching in graphene nanogaps in ambient conditions. For PCM devices with electrode separations less than 5 nm we find the resistance switching to be fully dominated by the formation of carbon filaments. While the actual mechanisms that we propose (carbon filamentation) need further unambiguous proof, nonetheless, our results point towards a key scaling limit to using such electrodes.

Thus, electric-field driven resistance switching in graphene nanogaps constrains the operational voltages possible in such devices. We find that at room temperature, switching can occur at $V_{th} < 0.4$ V, which, for example is the typical operating voltage for single-molecule devices. The noise observed in graphene-based single-molecule transistors at room temperature is likely to be the result of rearrangement of atoms and bonds at the edges of the electrodes. The fact that this noise is not observed at cryogenic temperatures agrees with previous observations that the resistance switching process is thermally assisted. Our results highlight the importance of gaining better knowledge of the edge chemistry in graphene nanogaps. These initial findings need further investigation by research groups specializing in techniques such as atomic-scale imaging to verify the nature of these atomic chains, as well as the influence of the actual material in the gap on the formation of these chains.

Although the potential formation of graphene filaments poses challenges to the development of graphene-based nanoelectrodes, it also offers exciting opportunities to study charge transport in atomic carbon chains. The formation of cumulene and polyyne chains have been observed using transmission electron microscopy[34]. If these structures could be controllably formed between graphene nanoelectrodes, they could serve as a test bed for the observation of a plethora of transport phenomena predicted in atomic chains[28,29,35]